\documentclass[12pt]{article}
\usepackage{amssymb}
\usepackage{amsmath}
\usepackage{graphics}
\usepackage{epsfig}
\usepackage{harvard}
\usepackage[hang,small,tight]{subfigure}
\usepackage{xspace,colortbl}
\usepackage{bm}
\usepackage{graphicx}
\citationstyle{dcu}

\newcommand{\blind}{0}

\makeatletter

\@addtoreset{equation}{section}

\makeatother

\makeatletter
 
  \@addtoreset{figure}{section}
\makeatother

\makeatletter
 
  \@addtoreset{table}{section}
\makeatother
\def\reff#1{(\ref{#1})}
\def\sobre#1#2{\lower 1ex \hbox{ $#1 \atop #2 $ } }

\begin{document}

\def\E{{\mathbb E}}
\def\P{{\mathbb P}}
\def\R{{\mathbb R}}
\def\Z{{\mathbb Z}}
\def\V{{\mathbb V}}
\def\N{{\mathbb N}}
\def\NN{{\bf N}}
\def\X{{\cal X}}
\def\supp{{\rm Supp}\,}
\def\XX{{\bf X}}
\def\Y{{\bf Y}}
\def\G{{\cal G}}
\def\T{{\cal T}}
\def\cC{{\cal C}}
\def\C{{\bf C}}
\def\D{{\bf D}}
\def\U{{\bf U}}
\def\K{{\bf K}}
\def\H{{\bf H}}
\def\n{{\bf n}}
\def\m{{\bf m}}
\def\b{{\bf b}}
\def\g{{\bf g}}
\def\sqr{\vcenter{
         \hrule height.1mm
         \hbox{\vrule width.1mm height2.2mm\kern2.18mm\vrule width.1mm}
         \hrule height.1mm}}                  % This is a slimmer sqr.
\def\square{\ifmmode\sqr\else{$\sqr$}\fi}
\def\one{{\bf 1}\hskip-.5mm}
\def\liml{\lim_{L\to\infty}}
\def\given{\ \vert \ }
\def\ze{{\zeta}}
\def\be{{\beta}}
\def\de{{\delta}}
\def\la{{\lambda}}
\def\ga{{\gamma}}
\def\th{{\theta}}
\def\proof{\noindent{\bf Proof. }}
\def\rate{{e^{- \beta|\ga|}}}
\def\A{{\bf A}}
\def\B{{\bf B}}
\def\C{{\bf C}}
\def\D{{\bf D}}
\def\bb{{\bf b}}
\def\bd{{\bf d}}
\def\bm{{\bf m}}
\def\lnt{{\Lambda^N}}
\def\S{{\mathcal{S}}}
\def\basis{{\rm Basis}\,}
\def\life{{\rm Life}\,}
\def\birth{{\rm Birth}\,}
\def\death{{\rm Death}\,}
\def\flag{{\rm Flag}\,}
\def\color{{\rm Color}\,}
\def\type{{\rm Type}\,}
\def\Cov{{\rm Cov}\,}
\def\btheta{{\boldsymbol{\theta}}}
\def\bphi{{\boldsymbol{\phi}}}
\def\bmu{{\boldsymbol{\mu}}}
\def\bbeta{{\boldsymbol{\beta}}}
\def\bEta{{\boldsymbol{\eta}}}
\def\bsigma{{\boldsymbol{\sigma}}}
\def\btheta{{\boldsymbol{\theta}}}
\def\bTheta{{\boldsymbol{\Theta}}}
\def\bSigma{{\boldsymbol{\Sigma}}}
\def\bvep{{\boldsymbol{\varepsilon}}}

\if0\blind
{
  \title{\bf A Hierarchical Model for Aggregated Functional Data}
  \author{Ronaldo Dias, Nancy L. Garcia \thanks{Corresponding
      author. E-mail addresses:
      { \tt dias@ime.unicamp.br}, {\tt nancy@ime.unicamp.br} and
      {\tt alex@im.ufrj.br}. This work was partially funded by CNPq grants 301530/2007-6, 301542/2007-4 and
475504/2008-9. The authors are grateful to NUMEC/USP, {\em N\'ucleo de
Modelagem Estoc\'astica e Complexidade} of the University of S\~ao
Paulo, for its hospitality.}\hspace{.2cm}\\
    Department of Statistics, University of Campinas (UNICAMP), Brazil\\
    and \\
     Alexandra M. Schmidt\\
    Department of Statistics, IM-UFRJ, Brazil}
  \maketitle
} \fi

\if1\blind
{
  \bigskip
  \bigskip
  \bigskip
  \begin{center}
    {\LARGE\bf A Hierarchical Model for Aggregated Functional Data}
\end{center}
  \medskip
} \fi

\bigskip
\begin{abstract}
  In many areas of science one aims to estimate latent sub-population
  mean curves based only on observations of aggregated population
  curves. By aggregated curves we mean linear combination of
  functional data that cannot be observed individually. We assume that
  several aggregated curves with linear independent coefficients are
  available. More specifically, we assume each aggregated curve is an
  independent partial realization of a Gaussian process with mean
  modeled through a weighted linear combination of the disaggregated
  curves. We model the mean of the Gaussian processes as a smooth
  function approximated by a function belonging to a finite
  dimensional space ${\cal H}_K$ which is spanned by $K$ B-splines
  basis functions. We explore two different specifications of the
  covariance function of the Gaussian process: one that assumes a
  constant variance across the domain of the process, and a more
  general variance structure which is itself modelled as a smooth
  function, providing a nonstationary covariance function. Inference
  procedure is performed following the Bayesian paradigm allowing
  experts' opinion to be considered when estimating the disaggregated
  curves. Moreover, it naturally provides the uncertainty associated
  with the parameters estimates and fitted values. Our model is
  suitable for a wide range of applications. We concentrate on two
  different real examples: calibration problem for NIR spectroscopy
  data and an analysis of distribution of energy among different type
  of consumers.
\end{abstract}

\noindent%

\newpage
%\spacingset{1.45} % DON'T change the spacing!

\section{Introduction}

The problem we address is the estimation of latent sub-population mean
and covariance curves when only populational aggregated data is available. By
aggregated data we mean that each sample consists of linear
combinations of functional data that cannot be observed individually
for each sub-population.

Certainly, there are many methods of fitting curves to data. A
collection of techniques known as nonparametric regression, for
example, allows great flexibility in the possible form of the
regression curve $\alpha$. In particular, it assumes no parametric form for
it. In fact, a nonparametric regression model only makes the assumption
that the regression curve belongs to some infinite collection of
curves. Consequently, in order to propose a nonparametric model one
may just need to choose an appropriate space of functions where he/she
believes that the regression curve lies. This choice, usually, is
motivated by the degree of smoothness of $\alpha$. Then, one uses the
data to determine an element of this function space that can represent
the unknown regression curve. Consequently, nonparametric techniques
rely more heavily on the data for information about $\alpha$ than their
parametric counterparts. Also, this flexibility on the form of the
curve allows one  to incorporate prior information.  The literature on nonparametric
regression is vast, for the interested reader we refer to the book of
\citeasnoun{euba:1999}.

The set up for  nonparametric regression assumes that an
unknown function $\alpha$ of one or more variables and a set of
measurements $y_1, \ldots, y_n$ are such that:
\[y_{i}={\cal L}_{i}\alpha + \varepsilon_{i}, \]
where ${\cal L}_{1},\ldots,{\cal L}_{n}$ are linear functionals defined on some
linear space $\cal H$ containing $\alpha$, and
$\varepsilon_{1},\ldots,\varepsilon_{n}$ are measurement errors usually
assumed to be independent, with common, zero mean normal distributions
with unknown variance $\sigma^{2}$. Typically, the ${\cal L}_{i}$ will be
point evaluations of the function $\alpha$. That is, ${\cal
  L}_{i}g=g(x_i)$ and $y_i=y(x_i)$, where $x_i$ are the explanatory
variables for $i=1,\ldots,n$.

The problem we address here is more general. We have several unknown
functions $\alpha_{c}, c=1, \ldots, C$ of one or more variables and the set of
measurements are given by
\begin{equation}
\label{eq:ourmodel}
y_{i}= \sum_{c=1}^{C}{\cal L}_{ic}\alpha_{c} + \varepsilon_{i}
\end{equation}
where
${\cal L}_{ic},i = 1,\ldots, I, c=1,\ldots, C$ are the linear
functionals. The problem is to estimate the functions $\alpha_c, c=1,
\ldots, C$ based on the measurements $y_i, i=1, \ldots, I$. 

More specifically, in our model we assume each aggregated curve $y_i$
is an independent partial realization of a Gaussian process with mean
modeled through a weighted linear combination of the disaggregated
curves $\alpha_c$s. Following \citeasnoun{dias:garcia:martarelli:2009}
we model the mean of the Gaussian process as a smooth function
approximated by a function belonging to a finite dimensional space
${\cal H}_K$ which is spanned by $K$ B-splines basis functions.  This
is not the only choice, other basis could be used such as Fourier
expansion, wavelets, natural splines. See, for example,
\citeasnoun{silv:1986}, \citeasnoun{koop:ston:1991},
\citeasnoun{vida:1999}, \citeasnoun{dias:1998a} and
\citeasnoun{dias:2000}. 

In this work, differently from
\citeasnoun{dias:garcia:martarelli:2009}, we consider two different
structures for the covariance function of the Gaussian process
postulating models that impose positive definiteness condition on the
function. The first one assumes a uniform structure across the domain
of the process, the second one models the covariance itself as a
smooth function providing a nonstationary behaviour. Our model is
going to be applied for two practical situations. In both of them it
is reasonable to consider that the experts in the field have prior
information on the disaggregated curves. Therefore, in order to incorporate
this opinion,  inference procedure will be performed following the Bayesian
paradigm. As a by-product, we naturally obtain the uncertainty associated
with the parameters estimates, and fitted values.

This paper is organized as follows. Section \ref{sec:ex} presents two
motivating examples: calibration problem for NIR spectroscopy data
and an analysis of distribution of energy among different types of
consumers.  Section \ref{sec:model} describes our proposed hierarchical
model to estimate latent disaggregate curves when only aggregated
population observations are available.  Therein we also propose a
non-stationary covariance function allowing the variance of the
underlying process to smoothly change across the domain of the
function. Next section analyzes different sets of artificial data. The
aim is to check the ability of the model in recovering the true
disaggregated functions under different scenarios. Then Section
\ref{sec:realdata} discusses the analysis for the two motivating
examples described in subsections \ref{sec:nir} and
\ref{sec:traficload}. Finally, Section \ref{sec:conclusion} concludes.

\section{Motivating examples} \label{sec:ex} 

\subsection{Near-infrared (NIR) spectroscopy data: a calibration
problem \label{sec:nir} }

Analyzing materials to determine their chemical composition is a basic
tool of science. It is used in many applications such as food safety
testing, protein detection, pharmaceutical purposes, forensics, to
mention just a few.  This analysis can be done directly in the
laboratory by usually expensive and time-consuming techniques. An
alternative is to determine the chemical composition through NIR
spectroscopy which is a low cost technique, relatively simple to use
and provides adequate accuracy in many practical situations. Some
references with practical applications of NIR spectroscopy are
\citeasnoun{cand:mass:etal:1999}; \citeasnoun{rodriguez-saona:2001};
\citeasnoun{mara:catt:gian:2002}; \citeasnoun{tewa:mehr:iru:2003};
\citeasnoun{cozzo:flood:etal:2006};
\citeasnoun{scho:mohl:wint:reic:2006}; Saranwong and Kawano (2008a,
2008b); \nocite{sara:kawa:2008a} \nocite{sara:kawa:2008b}
\citeasnoun{wood:dow:don:2008}; \citeasnoun{boto:bow:lamo:tsa:2009};
and \citeasnoun{romia:berna:2010}. For introductory material on the
subject see Shenk and Westerhaus (1991), Brereton (2003), and Burns
and Ciurczac (2007).

When atoms or molecules absorb light, the energy input excites a
quantized structure to a higher energy level. The type of excitation
depends on the wavelength of the light. NIR spectroscopy technology is
based on the fact that each of the major chemical components of a
sample has near infrared red absorption properties in the region
700-2500 nm. An absorption spectrum is the absorption of light as a
function of wavelength. The NIR spectrum of a sample is the summation
of these absorption properties for each chemical sample resulting in a
continuous curve measured by modern scanning instruments at hundreds
of equally spaced wavelengths.  The information contained in this
curve can be used to predict the chemical composition of the
sample. The problem lies in extracting the relevant information from
possibly thousands of overlapping peaks. This can be accomplished by
applying the Beer-Lambert Law. The Beer-Lambert law is the linear
relationship between absorbance and concentration of absorbing
species.

\begin{figure}[!bpth]
\centering
\includegraphics[height=7cm,width=9cm]{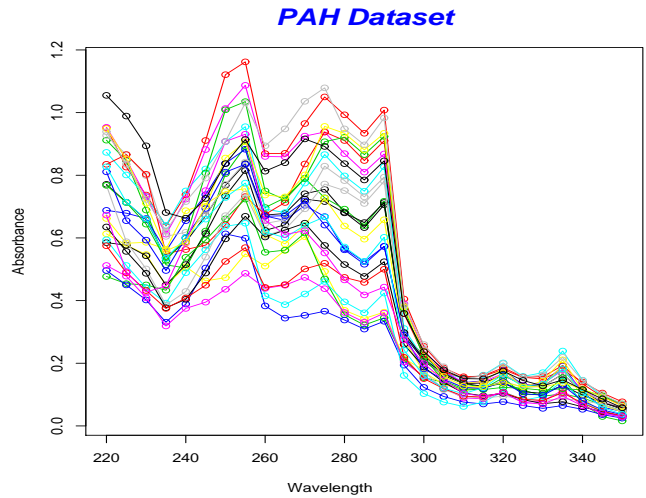}
\caption{Polyaromatic hydrocarbons spectra}
\label{PAHI_dados}
\end{figure}

Analysing a training set of different samples with distinct
compositions allow us to calibrate the analysis. Osborne, Fearn, and
Hindle (1993) described applications in food analysis and reviewed
some of the standard approaches to the calibration problem.
Multivariate calibration techniques are widely used in the literature
for this kind of problem. We propose a different approach by treating
this problem in the framework of functional data analysis. We regard
the response of interest as an aggregated continuous curve observed
only at a set of discrete points.  Therefore, having measurements of
the NIR spectrum for several chemical samples, with distinct
compositions, will allow us to estimate the typical curve for each
constituint of the sample. Figure \ref{PAHI_dados} shows the
absorbance curves measured for a dataset of 10 polyaromatic hydrocarbons
(PAH) obtained by Electronic Absorption Spectroscopy. The sample
consists of 25 chemical samples, each sample composed of varying
compositions of 10 different constituents (pyrene, acenaphthene,
anthracene, acenaphthylene, chrysene, benzanthracene, uoranthene,
uorene, naphthalene, phenanthracene). Each sample was submitted to 27
wavelengths (220nm--350nm). This dataset was presented by
\citeasnoun{brer:2003} to illustrate multivariate calibration
techniques. 

\subsection{Electric load \label{sec:traficload}}

The distribution of electric energy is done in several stages: first
substations provide energy for regions in the city. This energy
arrives at power transformers ({\it trafo}, an usual acronym for
transformer) that redistributes it to micro-regions. Each micro-region
is composed of different types of consumers, residential, commercial,
industrial, among others. For each type of consumer, there are peaks
of consumption at certain hours of the day. In Brazil, for example, it
is empirically known that residential consumers have a peak on energy
consumption between 6--8pm (due partially to the use of electric
showers) and commercial and industrial consumers have their peak
between 8am--6pm. To avoid overload, trafos have to be designed to
deal with the maximum load of the day. Ideally, the distribution of
electric energy should be done in such a way that there is a constant
load during the whole day, all days of the week, all over the year for
all power plants, substations and transformers. Therefore, to have a
more efficient and uniform distribution of electricity, it is
necessary to know the profile of the consumption for each type of
consumer.  For each type of consumer, this typical curve is called the
{\it typology}. The empirical evidence described before might be used
as prior information when modeling the typology for each type of
consumer. 

\begin{figure}[!htb]
\begin{center}
\includegraphics[scale=0.55]{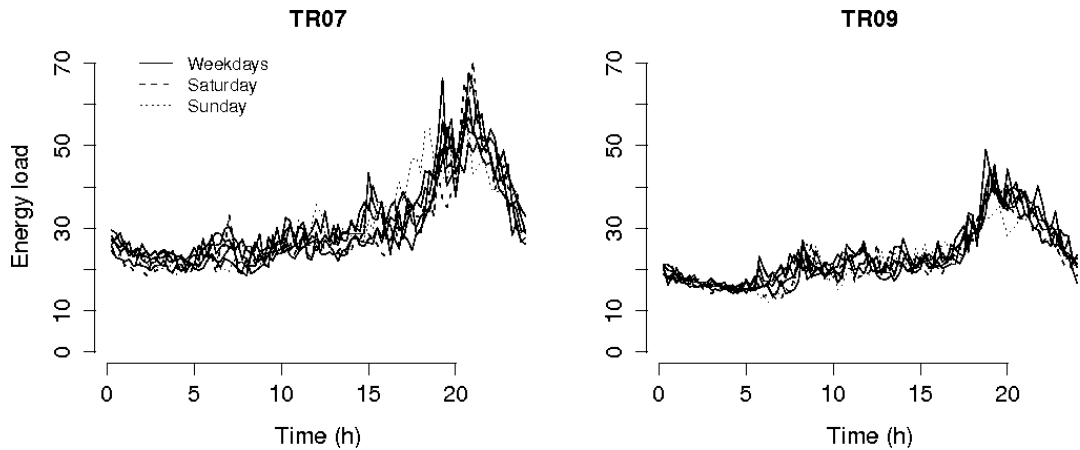}
\caption{Observed curves for two trafos.}
\label{fig:a1}
\end{center}
\end{figure}

From a practical point of view, it is very difficult and expensive to
obtain samples from individual consumers. Commonly, the data available
are aggregated data from power transformers (trafos). Typically each
trafo comprises around 50 consumers. Notice that this data is the sum
of all load demanded by the {\it market} (the number of consumers of
each type) of this trafo. Moreover, due to billing issues, the
market of each trafo is known. Therefore, having measurements of the
electric load for several different trafos, with distinct markets,
provides us with the information to estimate the individual curves for
each type of consumer. Figure \ref{fig:a1} shows the data from two trafos, that we analyse in Section \ref{sec:ex} 

%\vspace{1cm}

Both problems described above can be viewed as examples of samples
obtained from aggregated data; that is, the observed data might be
described as a linear combination of individual functional processes
and the aim is to estimate their individual mean and covariance
functions.

\section{Proposed Model \label{sec:model}}

Let $Y_{ij}(t)$ be the $j$-th replication of curve $i$ observed at
point $t$, $t \in [0,T_F]$. We decompose $Y_{ij}(\cdot)$ as the sum of
two components. The first one is described as a weighted sum of $C$
smooth curves, each representing the mean curve of a category  $c$
($c=1,\cdots,C$). For example, in the electric load example, each
disaggregated curve represents the typical curve of consumer type
$c$.
 The second component represents measurement error, described by a zero
mean Gaussian process with some covariance function. More
specifically, we assume
\begin{equation}
\label{model1} Y_{ij}(t)=\sum_{c=1}^{C} r_{ic} \alpha_{c}(t) +
\sum_{c=1}^C {\boldmath{\textbf{$\varepsilon$}}}_{ijc}(t), \quad
i=1, \ldots, I,\, \, j=1, \ldots, J,
\end{equation}
\noindent where $\alpha_1(t), \ldots, \alpha_C(t)$ are the mean curves
related to category $c=1,2,\cdots,C$, respectively. The $r_{ic}$s are assumed known and are
related to the problem being investigated. These will be discussed
in detail in Section \ref{sec:realdata}. We assume
${\boldmath{\textbf{$\varepsilon$}}}_{ijc}(\cdot)$ follows independent
zero mean Gaussian processes with covariance function given by
$Z_{ic}(t,s)$, for $t,s \in [0,T_F]$.

In general, the required degree of smoothness depends on the problem
under study. However, it is common to require that the functions
$\alpha_c$ belong to the Sobolev space ${\cal H}_{2}^{2}=\{f:[x_a,x_b]
\rightarrow \R, \sum_{j=0}^{2} \int (f^{(j)})^2 < \infty\} $.  To
consider the class ${\cal H}_{2}^{2}$ as the set of possible mean
curves is natural and desirable for this particular situation since it
can be well approximated by a finite-dimensional approximating space
generated by cubic B-splines, see \citeasnoun{boor:1978}. Therefore, the
second level of hierarchy expands the mean curves, $\alpha_c(\cdot)$,
as a linear combination of B-spline basis. We assume there exists a
positive integer $K$ and a knot sequence $\xi=(\xi_1,\cdots,\xi_K)$
such that
\begin{equation}
\label{eq:e3} \alpha_c(t)\,=\,\sum_{k=1}^{K} \beta_{ck} B_{k}(t),
\end{equation}
\noindent where $B_k(t)$, $k=1, \ldots, K$ are cubic
B-splines. Consider the function is evaluated at $T$ points, with $t_1
< t_2 < \cdots < t_T$, following equation (\ref{eq:e3}), we can write
\begin{eqnarray}\label{eq:ee3}
\left( \begin{array}{c} \alpha_{c}(t_1) \\ \vdots \\
    \alpha_c(t_T)\end{array} \right)\, = \,  \left( \begin{array}{c c c} B_1(t_1) & \cdots & B_K(t_1) \\ \vdots & & \vdots \\ B_1(t_T)& \cdots & B_K(t_T) \end{array} \right)\,
\left( \begin{array}{c} \beta_{c1} \\ \vdots \\
\beta_{cK}\end{array} \right)
\end{eqnarray}
for $c=1,2,\cdots,C$. 

Notice that the design matrix in equation \reff{eq:ee3} does not
depend on the category $c$ since we are using the same number of
basis and same knot allocation for all categories. Moreover, in this
model the coefficients do not depend on the sampled points and all
$N$ points of all aggregated curves can be used to estimate the same
$C \times K$ coefficients. Therefore, following equation
\reff{model1} and the discussion above,  we have the following
linear model
\begin{equation}\label{eq:e4}
Y_{ij}(t)=\sum_{c=1}^{C} \sum_{k=1}^{K} r_{ic} \beta_{ck} B_{k}(t)+
{\boldmath{\textbf{$\varepsilon$}}}_{ij}(t),
\end{equation}
where ${\boldmath{\textbf{$\varepsilon$}}}_{ij}(t) = \sum_{c=1}^C
{\boldmath{\textbf{$\varepsilon$}}}_{ijc}(t)$, because of the
independence assumption of $\epsilon_{ijc}(\cdot)$ for $i=1,\cdots,I$,
$j=1,\cdots,J$, and $c=1,\cdots,C$. We now discuss in detail the
covariance structure among the $\epsilon_{ij}(\cdot)$s.

\subsection{Covariance structure of the measurement error \label{sec:covariance}}

The measurement error captures any structure left after adjusting
the data to the sum of the latent disaggregated curves $\alpha_c(.)$. As we
are estimating functions we assume the errors are correlated across the domain of $Y_{ij}(\cdot)$.

We expect the correlation between points $Y_{ij}(t)$ and
$Y_{ij}(s)$ to decay exponentially,
as $|t-s|$ increases. For each category $c$, we assign an exponential correlation function with
decay parameter $\phi_c>0$ for the Gaussian process associated to each $\epsilon_{ijc}(\cdot)$. Our main contribution lies on the
specification of the variance structure. For the $i$-th curve, let
$Z_{i}(t,s) = {\rm Cov
}({\boldmath{\textbf{$\varepsilon$}}}_{ij}(t),
{\boldmath{\textbf{$\varepsilon$}}}_{ij}(s))$, be the covariance
between points $t$ and $s$. Notice that we assume the same covariance
structure across replicates $j=1,\cdots, J$. We propose
the following general structure for $Z_i(t,s)$,
\begin{equation}
\label{eq:e3a} Z_{i}(t,s) = \sum_{c=1}^C C_{ic} \eta_c(t) \,
\eta_c(s) \exp(-\phi_c|t-s|),
\end{equation}
where $C_{ic}, c=1, \cdots, C$, $i=1, \cdots, I$ are known
constants. Like the constants $r_{ic}$ in equation \reff{eq:e4}, the $C_{ic}$s
assume values related to the problem being studied.  For every $i$, we
allow the variances to change with $t$, such that $Z_i(t,t)= \sum_{c=1}^C
C_{ic} \eta_c(t)^2$. More generally, the covariance function is
allowed to change along the domain of the function. Because of the
product $\eta_c(t) \, \eta_c(s)$ in equation (\ref{eq:e3a}) we do not
need to impose any particular restriction on the $\eta_c(\cdot)$s to
guarantee that we have a valid covariance function. It is worth noting
that this covariance function might assume negative values, depending
solely on the function $\eta_c(\cdot)$.

We consider three different models for the components $\eta_c(\cdot)$s:
\begin{itemize}
\item[(a)] {\bf Uniformly homogeneous case}: In this case we assume 
$$
\eta_c(t) = \sigma, \, \, \forall \, t \quad \mbox{ and } \quad \phi_c=\phi, 
$$
implying that  $Z_{i}(t,s) =
(\sum_{c=1}^C C_{ic}) \sigma^2 \exp(-\phi|t-s|)$, for all $c = 1,
\ldots, C$.

\item[(b)] {\bf Homogeneous case}: Here we relax the assumption of
  common $\sigma^2$ and $\phi$ by assuming 
$$
\eta_c(t) = \sigma_c, \, \, \forall \, t,
$$
which leads to $Z_{i}(t,s) = \sum_{c=1}^C C_{ic} \sigma_c^2 \exp(-\phi_c|t-s|)$, i.e.

\item[(c)] {\bf Heterogenous case}: The more general case expands
  $\eta_c(\cdot)$ in B-splines basis functions, such that
\begin{equation}
\label{eq:e3b} \eta_c(t)\,=\,\sum_{\ell=1}^{L} \theta_{c\ell}
B_{\ell}(t),
\end{equation}
\noindent where $B_\ell(t)$, $\ell=1, \cdots, L$ are cubic B-splines
with (possibly) a different knot sequence from that for
$\alpha_c(\cdot)$, say $\xi_{\eta}$,  and the covariance function follows the general structure shown in Equation (\ref{eq:e3a}). 
\end{itemize}

\subsection{Likelihood function and Prior specification \label{sec:likel}}

Assume ${\bf y}$ represents the $IJT$-dimensional vector of
observations, with components ${\bf y}=(y_{11}(t_1),\cdots,\newline
y_{1J}(t_1),
\cdots,y_{11}(t_T),\cdots,y_{1J}(t_T),\cdots,y_{IJ}(t_T))$.
Considering the more general case, denote by
${\bTheta}$  the parameter vector
for the model. We will specify this vector for the three covariance
structures considered.  Notice that
for each $i=1,\cdots,I$, and $j=1,\cdots,J$, conditioned on the
parameter vector, ${\bf Y}_{ij}=(Y_{ij}(t_1),\cdots,Y_{ij}(t_T))'$
follows independent normal distributions such that%  \newline $Y_{ij}(t)
% \sim N\left( \sum_{c=1}^C \sum_{k=1}^K r_{ic} \beta_{ck} B_k(t),
%   Z_i(t,t)\right)$, then we can write
\begin{equation}
  \label{eq:n1}
   {\bf Y}_{ij} \sim N_T(X_i {\bbeta}, Z_i),
\end{equation}
where $X_i$ are $T \times CK$ matrices given by
$$
X_i = \left( \begin{array}{ccccccc}
               r_{i1} B_1(t_1) & \ldots &  r_{i1}B_K(t_1) & \ldots &
               r_{iC}B_1(t_1)& \ldots & r_{iC} B_K(t_1) \\
               \vdots & \ldots & \vdots &  \ldots & \vdots & \ldots &
               \vdots \\
               r_{i1} B_1(t_T) & \ldots &  r_{i1}B_K(t_T) & \ldots &
               r_{iC}B_1(t_T)& \ldots & r_{iC} B_K(t_T)
               \end{array}
         \right),
$$
${\bbeta}=({\bbeta}_1,\cdots,{\bbeta}_C)'$, with
${\bbeta}_c=(\beta_{c1},\cdots,\beta_{cK})'$, is the $CK$
dimensional vector of coefficients, and $Z_i = Z_i(\bTheta)$ are covariance
matrices of order $T$ with elements given by Equation \reff{eq:e3a}.
Therefore, based on the observed vector ${\bf y}$, the likelihood
function for ${\bTheta}$ can be written as
\begin{equation}
  \label{eq:n2}
   L({\bTheta};{\bf y}) \propto \prod_{i=1}^{I} \, \prod_{j=1}^{J}  |Z_i|^{-1/2}
\exp \left\{ -\frac{1}{2} ({\bf y}_{ij} - X_i {\bbeta})' Z_i^{-1} ({\bf
y}_{ij} - X_i {\bbeta}) \right\}.
\end{equation}
As our inference procedure follows the Bayesian paradigm, we now
specify the prior distribution of the parameter vector
${\bTheta}$ depending on the covariance structure.

\paragraph{Prior specification}
We assume prior independence among the components of the parameter vector ${\bTheta}$.
In particular, for the coefficients $\bbeta_c$ we assign 
$K$-dimensional multivariate normal distributions with known mean vector
${\bf b}_c$ and covariance matrice $\Omega_c$, $c=1,2,\cdots,C$. 
In Section \ref{sec:realdata} we assume a
zero mean prior for ${\bbeta}_c$. However, experts in the field
of interest might provide useful information about the shape of each
function, and this can be induced through the mean of the prior
distributions of the respective ${\bbeta}_c$s. For the covariance
matrices $\Omega_c$ we assume diagonal matrices, with the
diagonal elements fixed at some large value to let the observed data
drive the inference procedure. The prior specification of the parameters in the covariance function is related to the choice of the covariance structure proposed in Section \ref{sec:covariance}.

\begin{description}
\item [(a)] {\bf Uniformly homogeneous case:}
in this case the parameter vector is defined as $\bTheta^U=({\bbeta},\sigma^2,\phi)$.
For $\sigma^2$ we assume an inverse gamma
distribution with shape parameter $d$ and rate parameter $l$. For
$\phi$, we assign a gamma prior distribution with
  shape parameter $p$ and rate parameter $q$ fixed at some reasonable
  value.  For example, we can use the idea
of practical range. The mean of the prior, $p/q$ can be fixed such
that at a reasonable distance, the correlation is close to zero, say
0.05. More specifically, we fix the mean at the value that solves
$0.05=\exp(-\phi^* dist)$, where $\phi^*$ is the prior mean guess we
need, and $dist$ is a fixed distance.

\item [(b)] {\bf Homogeneous case:} for the homogenous covariance structure we define the parameter vector as $\bTheta^H=(\bbeta,\sigma^2_1,\sigma^2_2,\cdots,\sigma^2_C,\phi_1,\phi_2,\cdots,\phi_C)$.
We suggest independent inverse gamma prior distributions, each with parameters $(d_c,l_c)$,  for each $\sigma_c^2$.
We also assume prior independence among the decay parameters $\phi_c$ of the exponential correlation function; and each one is assumed to follow a gamma prior distribution with parameters $p_c$ and $q_c$, $c=1,\cdots, C$, with $p_c$ and $q_c$ fixed at some reasonable values.  The same idea of practical range discussed in the uniformly homogenous case can be used here.

\item [(c)]{\bf Heterogeneous case:}
the parameter vector to be estimated is $\Theta^{NH}=(\bbeta,\btheta_1,\cdots,\btheta_C,\phi_1,\cdots,\phi_C)$. We assume independent $K$-dimensional multivariate normal prior  distributions for the coefficients $\btheta_c$, each with known mean vector ${\bf d}_c$, and covariance matrix $\Lambda_c$. 
Like in the mean values of $\bbeta$, the  ${\bf d}_c$s can be obtained by experts in
  the field. However, it might be more challenging to elicitate these values as they are in the covariance structure of the process.  For the covariance
  matrices $\Lambda_c$ we assume diagonal matrices, with
  the diagonal elements fixed at some reasonably large value to let the observed
  data drive the inference procedure.
\end{description}

\paragraph{Posterior distribution and inference procedure}
Following the Bayesian paradigm, the posterior distribution,
$p(\bTheta \mid {\bf y})$, is proportional to the likelihood
function times the prior distribution of $\Theta$.

The resultant posterior distributions under all different covariance functions do not have 
closed forms. We use Markov chain Monte Carlo (MCMC) methods,
specifically, the Gibbs sampler with some steps of the Metropolis-Hastings (M-H)
algorithm to obtain samples from the target posterior distribution (see e.g.
\citeasnoun{game:lop:2006}). In particular, the 
full conditional posterior distributions of $\bbeta_c$ are normal distributions,
which are easy to sample from. Independent of the assumed covariance function, the  full conditional posterior distributions of each of the parameters involved in it do not result in known distributions.
For these parameters we make use of the Metropolis-Hastings algorithm with
log-normal proposals based on the current value of the chain, and some fixed variance, tuned to give reasonable acceptance rates. The MCMC
algorithm was implemented in {\tt R} \cite{Rproj},
and the codes are available from the authors upon request.

\subsection{Predictive inference \label{sec:predictive}}

From a Bayesian point of view, one can obtain the posterior predictive distribution of the function $Y_i(\cdot)$ at unobserved values of the domain of the function, ${\bf Y}_i^*=(Y_{i}(t_1^*),\cdots,Y_{i}(t_L^*))$, for $t_l^* \in [0,T_F], \, l=1,\cdots,L$, through
\begin{eqnarray}
p({\bf y}_i^* \mid {\bf y})=\int_{{\bTheta}} p({\bf y}_i^*\mid
{\bf y},  {\bTheta}) p({\bTheta} \mid {\bf y}) d{\bTheta}. \label{eq:predictive}
\end{eqnarray}

The model assumes that samples $Y_i(\cdot)$ are being generated from the multivariate normal distribution, $N(X_i \bbeta,Z_i)$. From the theory on the multivariate normal distribution
\cite{anderson:1984}, it follows that the joint distribution of
 ${\bf Y}$ and ${\bf Y}_i^*$, conditioned on ${\bTheta}$, is given by
\begin{eqnarray}
\left(
\begin{array}{c}
{\bf Y}_i^* \\
{\bf Y}
\end{array} \bigg\vert{\bTheta} \right) \sim N \left( \left(
\begin{array}{c}
X_i^* \bbeta \\
X_i \bbeta
\end{array} \right);
\left(
\begin{array}{cc}
Z_i^* & Z_{i_{12}}'\\
Z_{i_{12}} & Z_i
\end{array} \right)
\right),
\end{eqnarray}
where $X_i^*$ is a L-dimensional vector with elements equal to the cubic B-splines at point $t_l^*$; $X_i$ is a vector comprising the cubic B-splines at the  observed points $t_t$; 
$Z_i^*$ is a covariance matrix of dimension $L$ and each of its element is
the covariance of the process between unobserved points. Each
line of the matrix $Z_{i_{12}}$, $T \times L$, represents the
covariance between the $i^{th}$ observed point and the $j^{th}$
unobserved one, $i=1,\cdots,T$ and $j=1,\cdots,L$. From the theory
of the multivariate normal distribution we have that
\begin{eqnarray}
{\bf Y}_i^* | {\bf y}_i,{\bTheta} \sim N_L
\left(X_i^* \bbeta+Z_{i_{12}}'{Z_i}^{-1} \left( {\bf
y}_i-X_i \bbeta \right);
Z_i^*-Z_{i_{12}}'Z_i^{-1}Z_{i_{12}} \right).
\label{preditivacondicional}
\end{eqnarray}
The integration in (\ref{eq:predictive}) does not have an analytical
solution, however approximations can be easily obtained through
Monte Carlo methods \cite{game:lop:2006}. For each sample
$s$, $s=1,\cdots, Q$, obtained from the MCMC algorithm, we can
obtain an approximation for (\ref{eq:predictive}), by sampling from the
distribution in (\ref{preditivacondicional}) and computing
\begin{eqnarray}
p({\bf y}_i^* | {\bf y}) \approx \frac{1}{Q}\sum_{s=1}^Q \,
p({\bf y}_i^* | {\bTheta}^s). \label{eq:predictivelik}
\end{eqnarray}
Once samples from the posterior distribution of $\bTheta$ are
available, realizations from the posterior predictive distribution can
be obtained by sampling from the distribution of ${\bf Y}_i^* \mid
{\bf y}, \bTheta^s$, with $\bTheta^s$ representing the $s$th sampled
parameter vector $\bTheta$.

\section{Analyzing artificial data sets \label{sec:artificial}}

Here we analyze six artificial sets of data to check the ability of
the model in estimating the disaggregated curves of interest when the
truth is known. All datasets assume $C=2$ population curves.  We
consider data are generated from
\begin{equation}
\label{ex1} Y_{ij}(t)= r_{i1} \alpha_1(t) + r_{i2} \alpha_2(t) +
\epsilon_{ij}(t), \quad i=1,2, 3,\, \, j=1, \cdots, J,
\end{equation}
where the true curves are given by
\begin{eqnarray*}
  \alpha_1(t) &=& 5 \exp\{-t\}  \sin(\pi\,t/2)  \cos(\pi\,t) \label{eq:alpha1.1} \\
\alpha_2(t) &=& 5 \exp\{-(t-0.2)\}  \cos(\pi\,t/2) \sin(\pi\,t),
\label{eq:alpha2.2}
\end{eqnarray*}
with $r_{11}= 1$, $r_{12}=4$, $r_{21}=4$, $r_{22}=1$, $r_{31} = 2.5$
and $r_{32}=2.5$. These curves were chosen because they have
interesting features to be captured by the model. 
We explore 6 different scenarios by assuming different specifications for the  covariance structures of $\epsilon_{ij}(\cdot)$:

\paragraph{Case 1: Uniformly homogeneous case} In this case we assume all
  $C_{ic}=1$, $\sigma^2 = 1$ and $\phi = 0.5$. We concetrated on the
  case where there are no replicates for the aggregated curves, such that $J=1$ and we obtained samples for $I=10$ and $I=30$. %The aim is to reproduce the is is inspired by the real data
%example where there are no replicates for the aggregated curves.

\paragraph{ Case 2: Homogeneous case}
Here we assume
\begin{equation}
\label{eq:e3a1} Z_{i}(t,s) =  C_{i1} \sigma_1^{2} \exp(-\phi_1|t-s|)
+ C_{i2} \sigma_{2}^2 \exp(-\phi_2|t-s|),
\end{equation}
and we fix the parameters at the following values: $\sigma_1^2 =
\sigma_2^2 = 1$, $\phi_1 = \phi_2 = 4$ and
$C_{11}= 1$, $C_{12}=1.3$, $C_{21}=1.4$, $C_{22}=1.3$, $C_{31} =
1.5$ and $C_{32}=1.5$.
Here we consider only the case $J=15$. 

\paragraph{Case 3: Heterogeneous case}
Here we assume
\begin{equation}
\label{eq:e3a2} Z_{i}(t,s) =  C_{i1} \eta_1(t) \eta_1(s)
\exp(-\phi_1|t-s|) + C_{i2} \eta_2(t) \eta_2(s) \exp(-\phi_2|t-s|)
\end{equation}
with $\eta_1$ and $\eta_2$ curves generated as linear combinations
of B-splines, $\phi_1 = \phi_2 = 4$ and $C_{11}= 1$, $C_{12}=1.3$,
$C_{21}=1.4$, $C_{22}=1.3$, $C_{31} = 1.5$ and $C_{32}=1.5$.
For the heterogeneous covariance
structure we fit the model considering $J=15$, $J=50$ and $J=150$.
This is to investigate the effect of the number of replicates on the
estimates of the parameters when a more flexible  covariance
structure is assumed.

For all datasets we assumed 14 B-splines basis with $K=10$ internal
knots. In the heterogeneous case, we assumed this same set of knots to
estimate $\eta_1(.)$ and $\eta_2(.)$.  We let the MCMC algorithm run
for 100,000 iterations, considered the first 5,000 as burn-in and kept
every $95$-th sample to avoid autocorrelation between the sampled
values. Convergence of the chains was checked through the use of two
chains starting from very different values.

\begin{figure}[!h]
\begin{center}
\includegraphics[width=13cm]{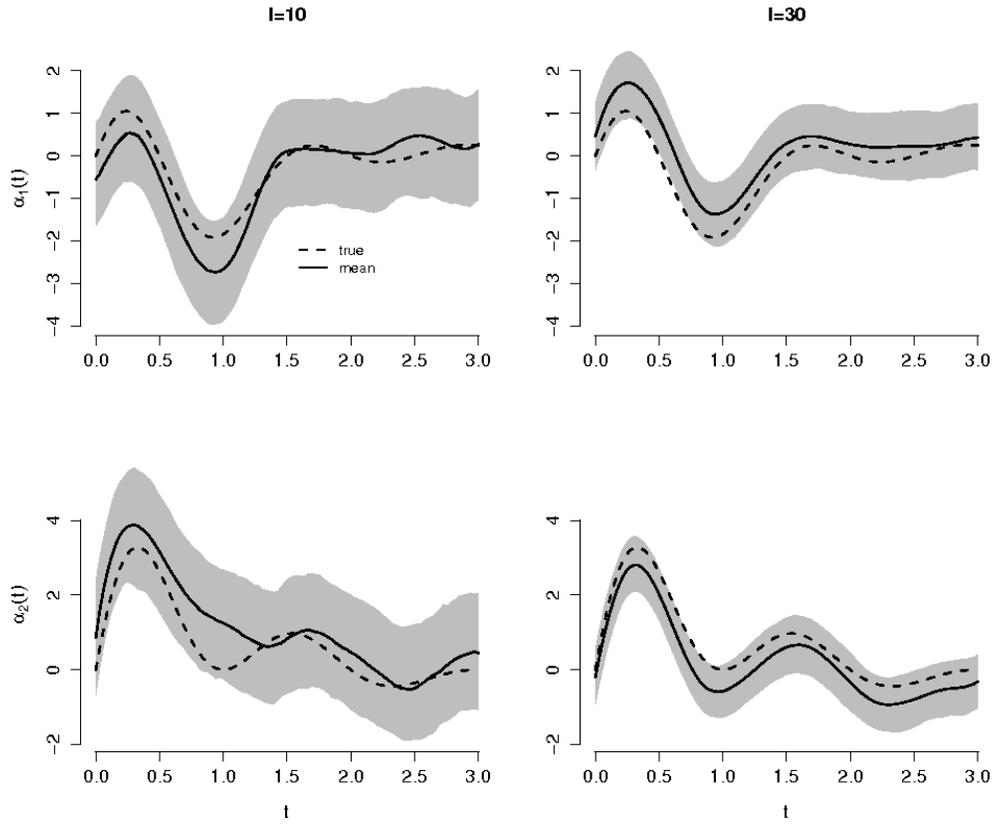}
\caption{Summary of the posterior distribution of the parameters for
  the uniformly homogeneous case with $J=1$, $I=10$ (first column) and
  $I=30$ (second column).  Posterior mean curves (solid lines) and
  limits (shaded area) of the 95\% posterior credible
  intervals. Dashed lines represent respective true
  values.} \label{fig:alphaUni}
\end{center}
\end{figure}

\begin{figure}[!h]
\begin{center}
\includegraphics[width=13cm]{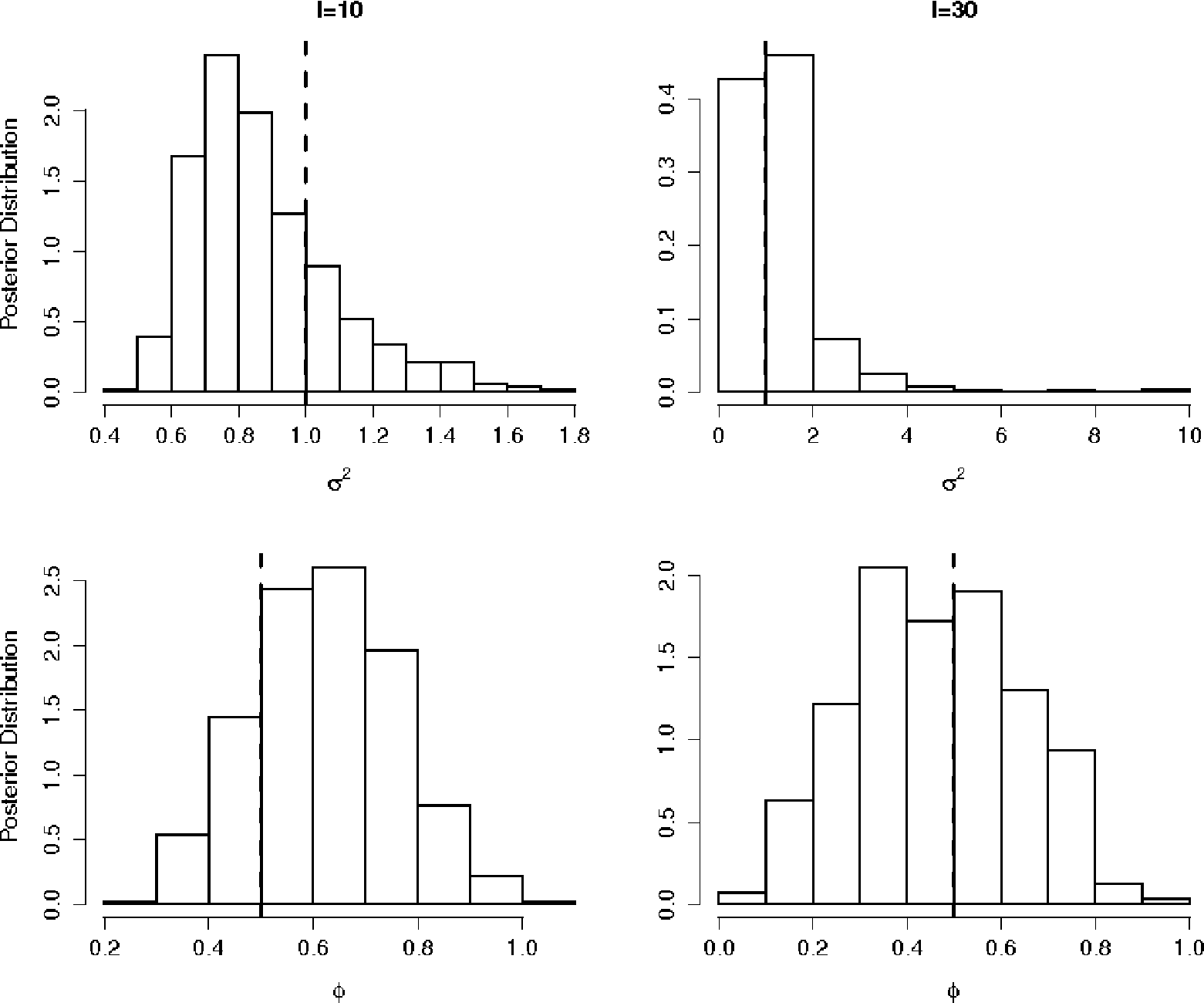}
\caption{Summary of the posterior distribution of the parameters for
  the uniformly homogeneous case with $J=1$ and $I=10$ (first row) and
  $I=30$ (second row). Posterior distribution of $\sigma^2$ and and
  $\phi$. Dashed lines represent respective true
  values.} \label{fig:histUni}
\end{center}
\end{figure}

For the uniformly homogeneous case we notice that inspite of the value of $I$ the posterior distribution of $\sigma^2$ and $\phi$ seem to recover the true values used to generated the data. On the other hand, the value of $I$ seems to have influence on the magnitude of the ranges of the 95\% posterior credible intervals  of the disaggregated curves $\alpha_c(\cdot)$ (Figures \ref{fig:alphaUni} and \ref{fig:histUni}).

\begin{figure}[!h]
\begin{center}
\includegraphics[width=15.0cm]{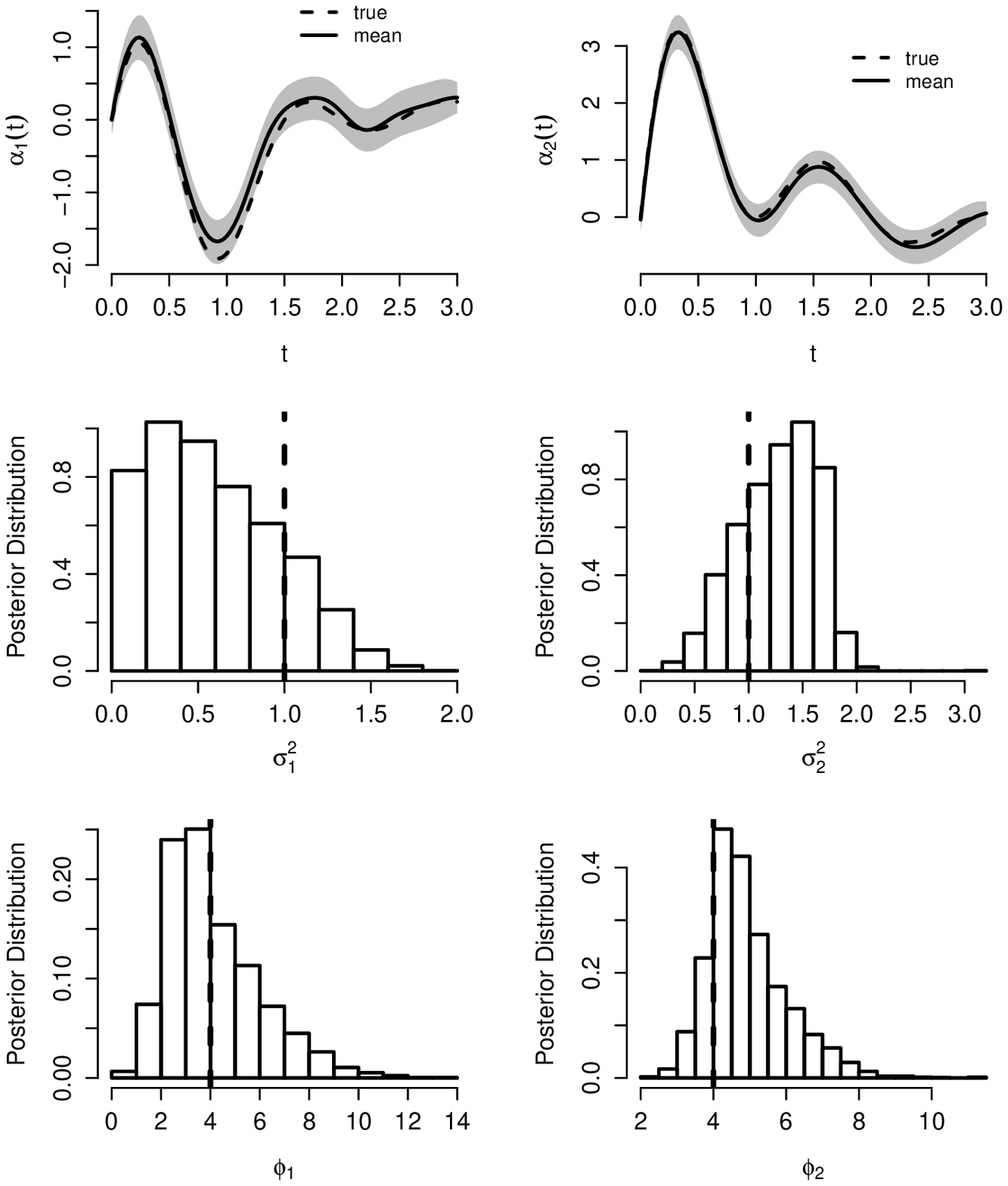}
\caption{Summary of the posterior distribution of the parameters for
the homogeneous case with $J=15$.  Posterior mean curves (solid
lines) and limits (shaded area) of the 95\% posterior credible
intervals (1st row), posterior distribution of $\sigma_1^2$ and
$\sigma_2^2$ (2nd row), and $\phi_1$ and $\phi_2$ (3rd row). Dashed
lines represent respective true values.} \label{fig:GRhom}
\end{center}
\end{figure}

For the homogeneous case, even with only $J=15$
replicates, the model is able to recover the true structure of the
disaggregated functions $\alpha_1(.)$ and $\alpha_2(.)$.
The posterior mean of both curves are very close to their respective
true values and the range of the 95\% posterior credible intervals
are relatively narrow.  The parameters in the covariance structure
are also well estimated (Figure \ref{fig:GRhom}).

\begin{figure}[!h]
\begin{center}
\includegraphics[width=15.0cm]{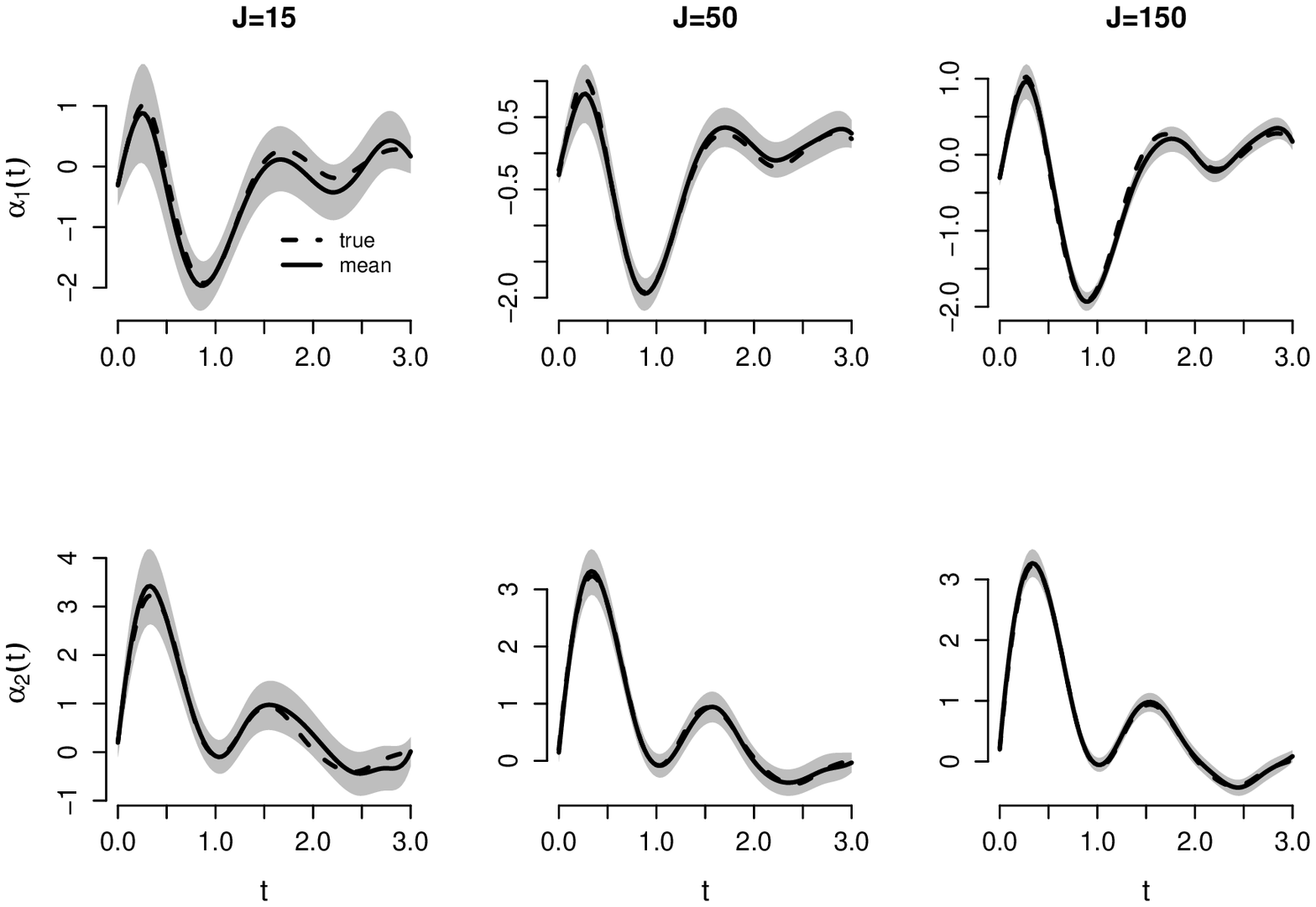}
\caption{Posterior mean curves (solid lines) and limits (shaded
  area) of the 95\% posterior credible intervals for (a)
  $\alpha_1(.)$, $J=15$, (b) $\alpha_2(.)$, $J=15$, (c) $\alpha_1(.)$,
  $J=50$, (d) $\alpha_2(.)$, $J=50$, (e) $\alpha_1(.)$, $J=150$, (f)
  $\alpha_2(.)$, $J=150$. In all panels the dashed line is the
  respective true curve.} \label{fig:mediasNP}
\end{center}
\end{figure}

\begin{figure}[!h]
\begin{center}
\includegraphics[width=15.0cm]{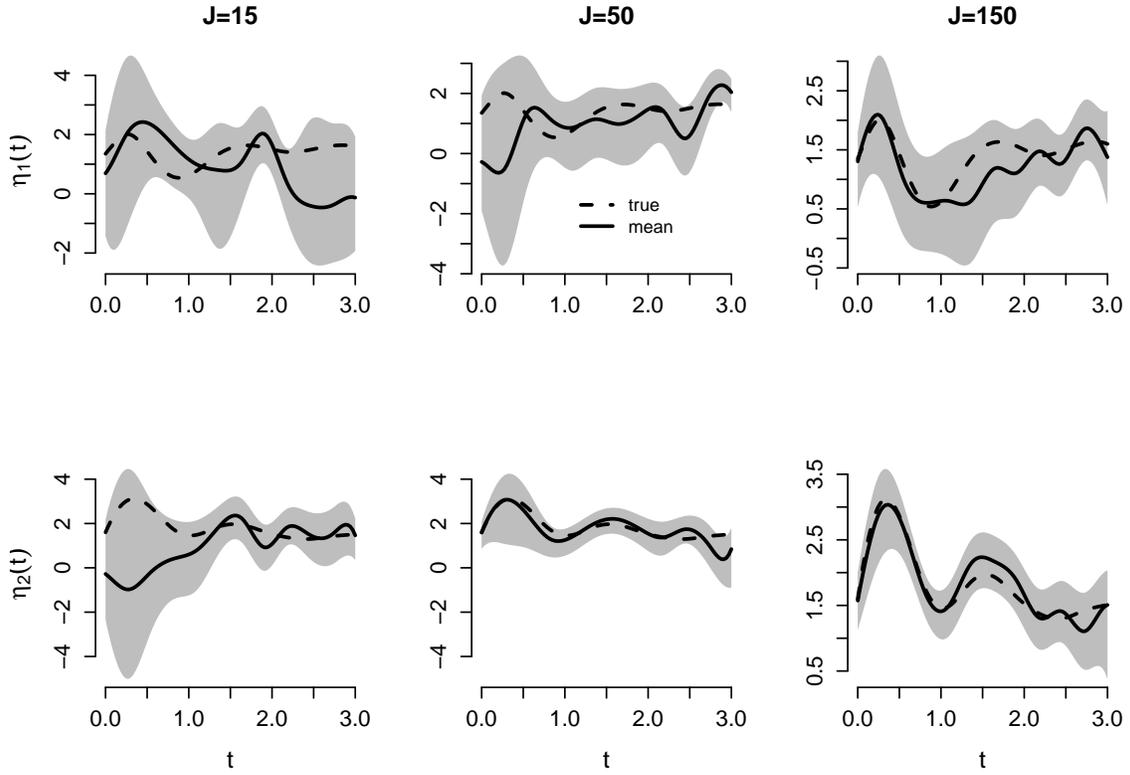}
\caption{Posterior mean (solid line) of the variance functions
  $\eta_1(.)$ and $\eta_2(.)$ (rows) and limits (shaded area) of the
  95\% posterior credible intervals for the heterogeneous case with
  $J=15$, $J=50$ and $J=150$ (columns). The dashed lines represent the
  true curves used to generate the data.}
\label{fig:etasNP}
\end{center}
\end{figure}

\begin{figure}[!htb]
\begin{center}
\includegraphics[width=8.0cm]{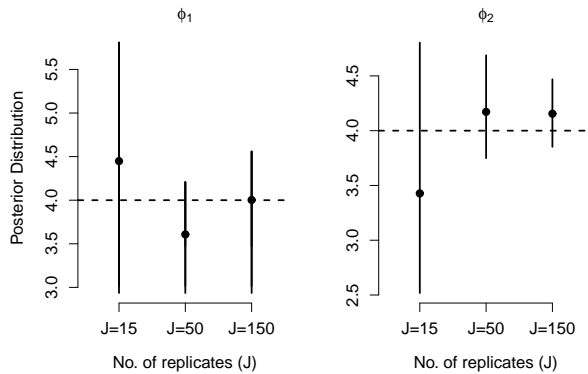}
\caption{Posterior summary (solid circle: posterior mean and lines represent the limits of the 95\% posterior credible intervals) of the decay parameters $\phi_1$ and $\phi_2$ in equation (\ref{eq:e3a}). Dashed line represents true value used to generate the data.  \label{fig:corr} }
\end{center}
\end{figure}

For the heterogenous case it is clear that the number of replicates
affect the range of the posterior credible intervals both for
$\alpha_c(.)$ and $\eta_c(.)$, $c=1,2$. Regardless of the value of $J$
the true values are recovered from the inference procedure, specially
for the disaggregated functions $\alpha_c(.)$ . The greater the number of
replicates the narrower the 95\% posterior credible intervals (Figures
\ref{fig:mediasNP} and \ref{fig:etasNP}). The true values of the decay
parameters $\phi_1$ and $\phi_2$ are also recovered from the inference
procedure, and similar to the results for $\alpha_c(.)$ and
$\eta_c(.)$, the magnitude of $J$ influences the range of the
posterior credible intervals (Figure \ref{fig:corr}). 

\clearpage

%\clearpage

\section{Applications \label{sec:realdata}}

\subsection{NIR Spectroscopy data - Polyaromatic hydrocarbons}

The Beer-Lambert law for $K$ constituents plus noise states that for
the $i$th chemical sample the measurement at wavelength $t$ is given
by
\begin{equation}
  \label{eq:modelchemo}
  Y_{i}(t)=\sum_{\ell=1}^{C} r_{\ell,i} \alpha_{\ell}(t) +
  \sum_{\ell=1}^{C} e_{\ell,i}(t), \quad
t \in [220,350], \quad i=1,\ldots,25.
\end{equation}
where $r_{\ell,i}$ is the concentration of the $\ell$ constituent in
the $i$th chemical sample , $\alpha_{\ell}(t)$ is the absorbance at
wavelength $t$ of the $\ell$th pure constituent and $e_{\ell,i}$ is a
random noise. 

Notice that the Beer-Lambert formula leads exactly to the functional
model of aggregated data given by Equation \reff{eq:ourmodel}.

In this example, we used 14 B-spline basis with 10 internal knots
equally spaced in the interval (220,350) to estimate the latent absorbance
curves of each constituint, $\alpha_c(\cdot)$. 

Since there are no replicates of the population curves available and
we have $C=10$ constituints, we chose to fit the model considering a
homogeneous covariance structure. In this case, we used a gamma prior
for $\phi_c$, and an inverse gamma with parameters 2 and 0.2 for
$\sigma_c^2$, $c=1, \ldots, 10$.  We let the MCMC algorithm run
for 100,000 iterations, considered the first 5,000 as burn-in and kept
every $100$-th sample to avoid autocorrelation between the sampled
values.

 \begin{figure}[!htb]
\begin{center}
   {\includegraphics[width=12.0cm]{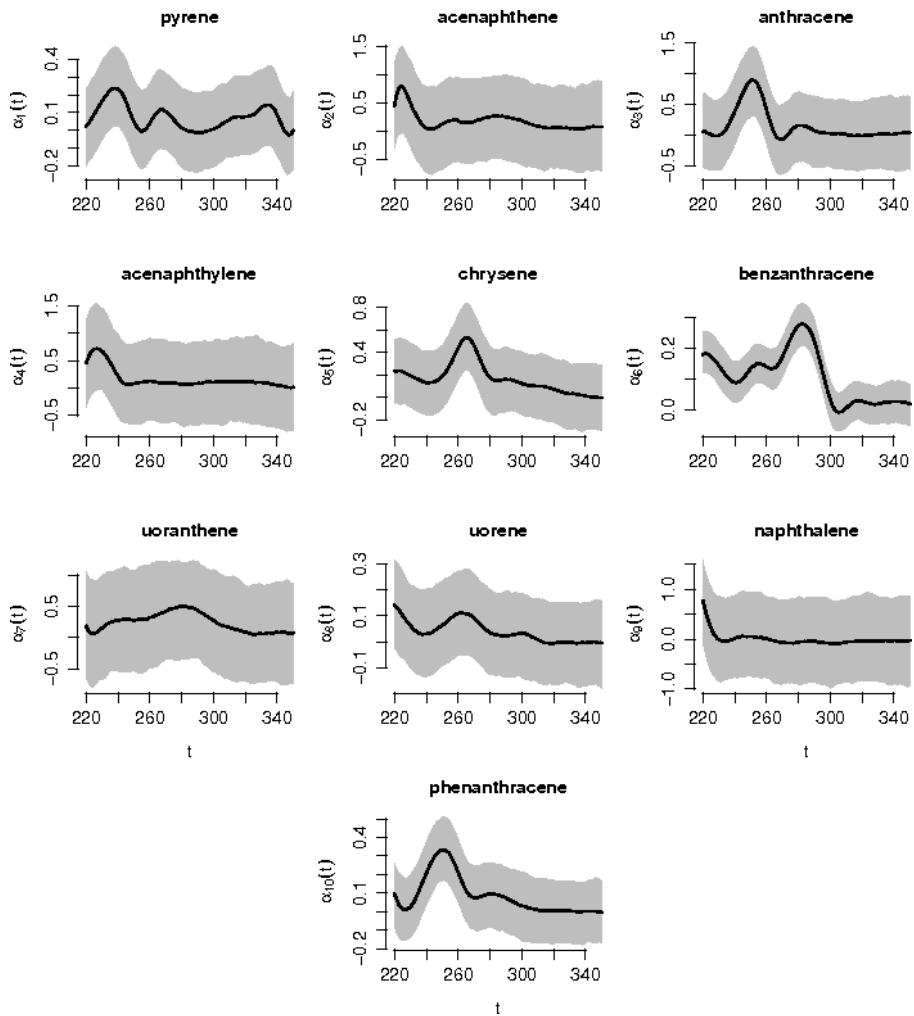}}
 \caption{Estimated absorbance curves for the constituints,
   for the PAH dataset under the uniformly homogeneous model.}\label{fig:abs}
 \end{center}
 \end{figure}

\begin{figure}[!htb]
\centerline{ {\includegraphics[width=7.0cm]{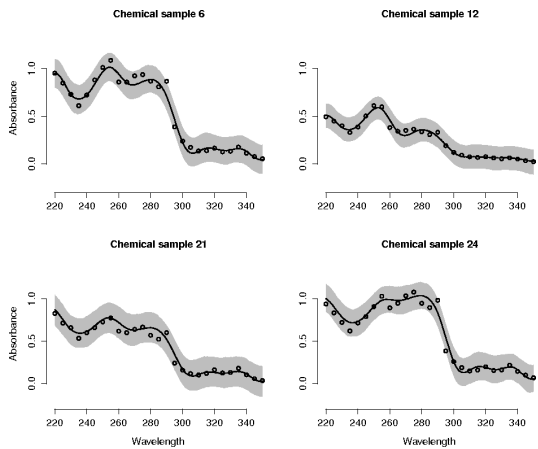}}}
\caption{Estimated curves for chemical samples 6, 12 ,21 and 24}
\label{fig:comp}
\end{figure}

\begin{figure}[!htb]
\centerline{ {\includegraphics[width=7.0cm]{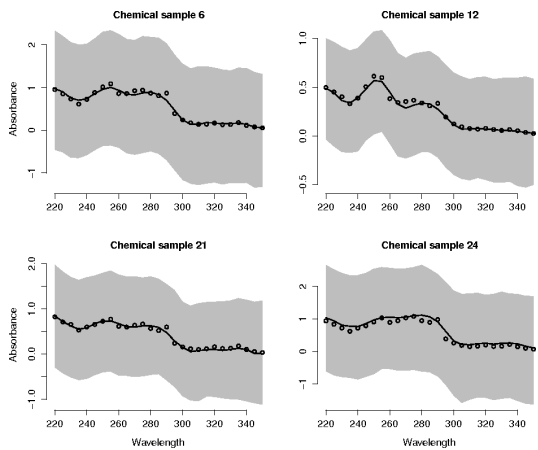}}}
\caption{Predictive 95\% confidence intervals for chemical samples 6, 12 ,21 and 24}
\label{fig:comppred}
\end{figure}

Figure \ref{fig:abs} shows the estimated individual curves for each of
the constituints. However, to see how good these estimates are we
should compare the aggregated observed curves with the weighted sum of
these estimates using the Beer-Lambert formula. These comparisons are
made in Figures \ref{fig:comp} and \ref{fig:comppred}. The confidence
bands for Figure \ref{fig:comp} are the quantiles weighted sum of the
estimates whereas  Figure \ref{fig:comppred} compares the data
obtained in some chemical samples with the predictive 95\% confidence
intervals. Both figures show that we have an excelent
fit for the data.

\subsection{Electric load data}

Here we analyze electric load data as described in Section \ref{sec:traficload}.
In Brazil, for security reasons, houses are loaded with energy tension
either equal to 127V or 220V. For this reason, they are classified as
monophasic (single phase/ 127V) or biphasic (two phases/220V). Usually, more modest residencies are monophasic 
suggesting that monophasic and biphasic consumers have different
typologies. We consider samples observed from $I=2$ trafos which
are denoted by TR07 and TR09.
The market for each trafo is small and
variable, consisting only of single phase and two phase
residential consumers.  The consumption of energy
during weekends is different than from weekdays (Figure \ref{fig:a1}). Therefore, we decide to analyse only the weekdays resulting that $J=5$ replicates.

Measurements from trafos TR07 and TR09 were stored at every 15
minutes, during 5 days of a particular week.  It is known that the electric load
of each trafo $i$ is equal to the sum of $N_i = N_{i1} + N_{i2}$
curves, where $N_{ic}$ is the number of consumers of type $c$ (monophasic or biphasic here), and
$(N_{i1}, N_{i2})$ is the market of trafo $i$. Table
\ref{tab:a1} presents the number of consumers (market) for each trafo.
\begin{table}[!h]
\begin{center}
\renewcommand{\arraystretch}{1}
%\caption{Market for trafos TR07 e
%  TR09.} \vspace{0.25cm}
\begin{tabular}{c|ccc} \hline \hline
Trafo & Single Phase & Two Phase   & Total \\ \hline
TR07      & 87                     & 5                    & 92    \\
TR09      & 25                     & 25                   & 50    \\
\hline  \hline
\end{tabular}
\end{center}
\caption{Distribution of the number of consumers (market) for trafos TR07 and TR09 in the electric load application.}
\label{tab:a1}
\end{table}

Following the general structure in equation (\ref{model1}), we model the traffic load of trafo $i$, at day $j$, observed at time $t$ as
\begin{equation}
  \label{eq:modelelectric}
  Y_{ij}(t)=\sum_{c=1}^{C} \sum_{n_c=1}^{N_{ic}} W_{cj n_c i}(t), \quad
t \in [0,24], \quad i=1,2, \quad j=1,\ldots,5,
\end{equation}
where $W_{c,j,n_c,i}(t)=\alpha_c(t)+\varepsilon_{c,j,n_c,i}(t)$. For this example the constants $r_{ic}$ and $C_{ic}$ (in equations (\ref{eq:e4}) and (\ref{eq:e3a})) coincide and are equal to $N_{ic}$.

\begin{figure}[!h]
\begin{center}
{\includegraphics[scale=0.58]{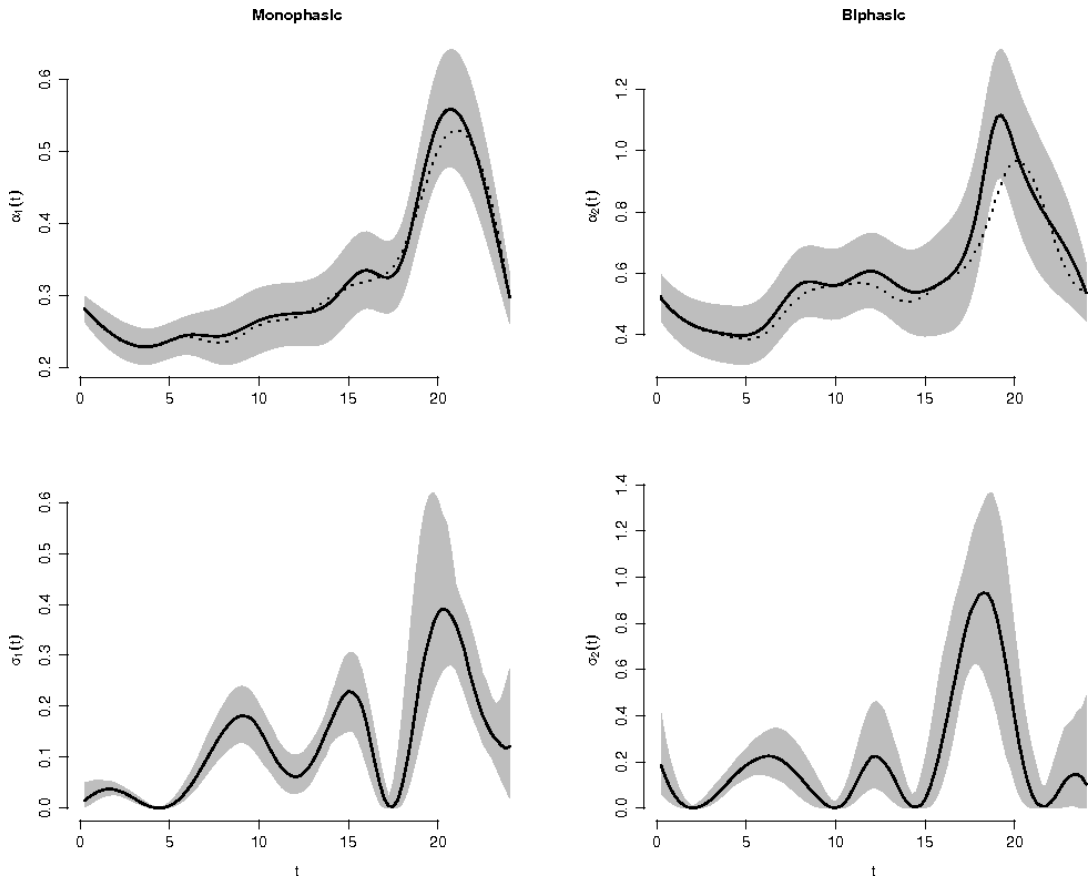}}
\caption{Posterior mean (solid line) and $95\%$ credible intervals
  (shaded area) for monophasic (single phase) and biphasic (two phase)
  consumers based on trafos TR07 and TR09 for the tipologies (first
  row) and variance (second row). The dotted lines in the first row show the
  estimate obtained under the frequentist approach proposed by Dias et
  al (2009). \label{fig:a4}}
  \end{center}
\end{figure}

We fitted the model assuming 14 B-spline basis, with $K=10$ internal knots located at the following points $\xi=\{4,6,8,10,12,14,16,18,19,20\}$. We fitted the model considering a heterogeneous covariance
structure. The prior specification of $\phi_c$ uses
the idea of practical range as discussed in Section \ref{sec:likel}.
{\em A priori}, we assume the correlation between measurements in a
day dies off, in average, after 3/4 of an hour. We assume a gamma
prior for $\phi_c$ with mean equals $3/0.75=4$, and variance 1.
Panels of figure \ref{fig:a4} show the posterior mean with
respective 95\% posterior credible intervals for the mean  and
variance curves for monophasic and
biphasic consumers. As expected  the single phase houses have a
smaller load than the two phase residencies. Both types have a peak
around 8pm, as this coincides with arriving home from work, taking
showers, etc. The basic difference is that for two phase residencies
there is an increase in the electric load from 8am to 12pm which is
not observed for the single phase consumers. 

%To check the goodness of fit, we estimate
%the electric load for each trafo through the estimated typologies
%and obtained a very good fit as can be seen from Figure
%\ref{fig:a5a}.
%There is a slight bias in the estimation for TR09
%between 5 and 8pm. This is caused probably by the fact that the
%number of single phase consumers for this trafo is significantly
%bigger than the two phase consumers.
% \begin{figure}[!htb]
% \begin{center}
% (a){\includegraphics[angle=270,width=7.0cm]{media1_reais_NP.ps}}
% (b){\includegraphics[angle=270,width=7.0cm]{media2_reais_NP.ps}} \\
% \caption{Posterior mean (solid line) and $95\%$ credible intervals
%   (dashed line) for (a) single phase and (b) two phase consumers
%   based on trafos TR07 and TR09. The dotted line shows the estimate
%   obtained under the  frequentist approach proposed by
%   Dias et al (2009) \label{fig:a4}}.
%   \end{center}
% \end{figure}

\begin{figure}[!htb]
\begin{center}
\includegraphics[scale=0.56]{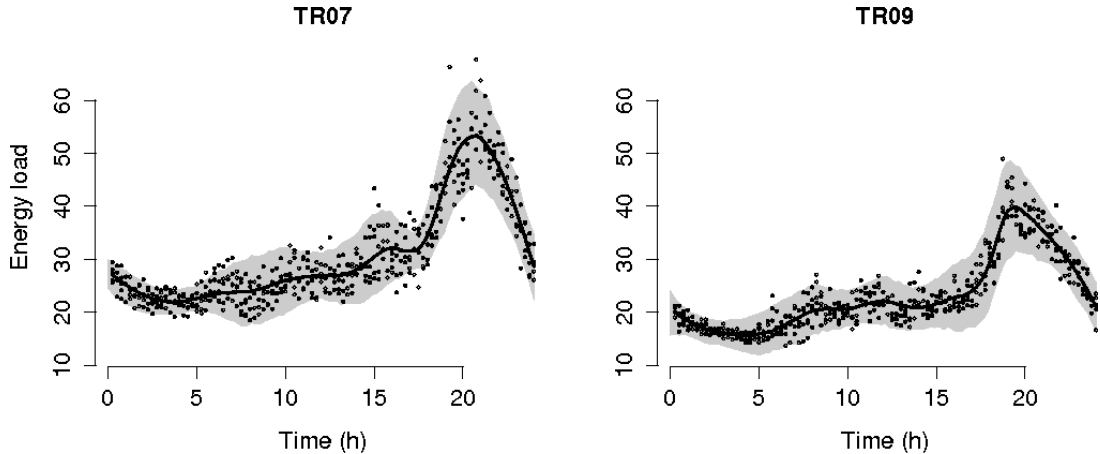}
\caption{Summary of the posterior predictive distribution (solid
  lines) and limits of the 0.025 and 0.975 quantile curves (shaded area)
  compared with the respective observed values (hollow circles), for trafos  TR07 and TR09. \label{fig:repelectric}}
  \end{center} 
\end{figure}

Figure \ref{fig:repelectric} compares the posterior predictive distribution with the observations. 
It is clear that the proposed model is capturing quite well the structure of the data, as most of the observations are within the limits of the 
95\% posterior predictive interval.

\section{Concluding remarks \label{sec:conclusion}}
In this work our attention was focused on the Bayesian estimation of
latent sub-population (disaggregated) mean and covariance curves when
we only have available observations of the population (aggregated)
curves.  Although our proposed models are an extension of the model
initially proposed by \citeasnoun{dias:garcia:martarelli:2009}, we
propose more flexible nonstationary covariance structures for the
Gaussian process by allowing the variance of the process to change
across the domain of the function. The general non-parametric case
naturally imposes the positive definiteness of the covariance function
and can be restricted to accomodate many different situations. We
believe our proposed covariance structure might also be applied in
different areas, e.g. geostatistics.
\newpage

There are several advantages of using the Bayesian
paradigm:
\begin{itemize}
\item  It might naturally incorporate prior information that is available to experts
  in the field; 
\item Estimates are obtained under a single framework;
\item and it naturally provides the uncertainties of the estimates of the
  latent sub-population curves, not only of the mean curves but also
  of the covariance curves.
\end{itemize}

To show the strength of our method, we analyzed two examples from
different areas in science: environmetrics and chemometrics. 
In both examples it is clear, from comparing the observed
aggregated curves with the weighted sum of the estimated latent ones,
that the proposed model provide extremely reasonable estimates.

\bibliography{myref}
\end{document}